\documentclass[%
 reprint,
showpacs,preprintnumbers,
 amsmath,amssymb,
 aps,
 prl,
 lengthcheck,%
]{revtex4}

\usepackage{ae,amssymb,amsmath}
\usepackage{graphicx}
\newcommand{\rb}{$^{87}$Rb}
\newcommand{\kq}{$^{41}$K}

\newcommand{\eref}[1]{Eq.~(\ref{#1})}
\newcommand{\fref}[1]{Fig.~\ref{#1}}

\newcommand{\bsym}{\boldsymbol}
\newcommand{\aeff}{a_{\rm eff}}
\newcommand{\mk}{m_{\rm K}}
\newcommand{\mrb}{m_{\rm Rb}}
\begin{document}
\title{Scattering in Mixed Dimensions with Ultracold Gases}

\author{G.~Lamporesi$^{1}$}
\author{J.~Catani$^{1,2}$}
\author{G.~Barontini$^{1}$}
\author{Y.~Nishida$^{3}$}
\author{M.~Inguscio$^{1,2}$}
\author{F.~Minardi$^{1,2}$}

\affiliation{$^1$LENS - European Laboratory for Nonlinear
  Spectroscopy and Dipartimento di Fisica, Universit\`a di Firenze,
  via N. Carrara 1, I-50019 Sesto Fiorentino, Italy\\
  $^2$INO-CNR, via G. Sansone 1, I-50019 Sesto Fiorentino, Italy\\
  $^3$ Center for Theoretical Physics, Massachusetts Institute of
  Technology, Cambridge, Massachusetts 02139, USA}

\begin{abstract}
  We experimentally investigate the mix-dimensional scattering
  occurring when the collisional partners live in different
  dimensions. We employ a binary mixture of ultracold atoms and
  exploit a species-selective 1D optical lattice to confine only one
  atomic species in 2D. By applying an external magnetic field in
  proximity of a Feshbach resonance, we adjust the free-space
  scattering length to observe a series of resonances in
  mixed dimensions. By monitoring 3-body inelastic losses, we measure
  the magnetic field values corresponding to the mix-dimensional scattering
  resonances and find a good agreement with the theoretical
  predictions based on simple energy considerations.

\end{abstract}


\date{\today}
\maketitle

Degenerate atomic gases have provided quantum systems with
unprecedented possibilities of manipulation and control, achieved
by combining magnetic and optical potentials as well as scattering
resonances. The capability to model and control  tightly confining
potentials sparked the experimental investigation on quantum
systems of reduced dimensionality, since particles can be forced
to occupy a single quantum level along specific directions.
Spectacular achievements, such as the observation of the BKT
crossover \cite{BKT} in 2D and the realization of Tonks-Girardeau
gases \cite{TG} in 1D, confirmed the importance of quantum gases
as testbench for fundamental low-energy physical phenomena.
Moreover, low dimensional ultracold atomic gases show further
peculiar scattering properties leading to the appearance of
confinement-induced resonances depending on the dimensionality of
the system \cite{PRL98-Olshanii, PRL03-Bergeman, PRA01-Petrov,
PRL05-Moritz}. Interestingly, while much of the work done so far
deals with well-defined dimensionality, systems composed of
interacting parts living in different dimensions have received
little attention and, besides recent theoretical analysis
\cite{PRA06-Massignan, PRL08-Nishida}, experimental investigation
is still lacking. Such mix-dimensional systems are relevant in
several physical domains, ranging from cosmology to condensed
matter physics. In brane theory, for example, particles and fields
are confined to the ordinary 3D space and interact with gravitons
that can propagate in extra dimensions \cite{PRL99-Randall}.\\
\begin{figure}[b]
  \centering
  \includegraphics[width=.9\columnwidth]{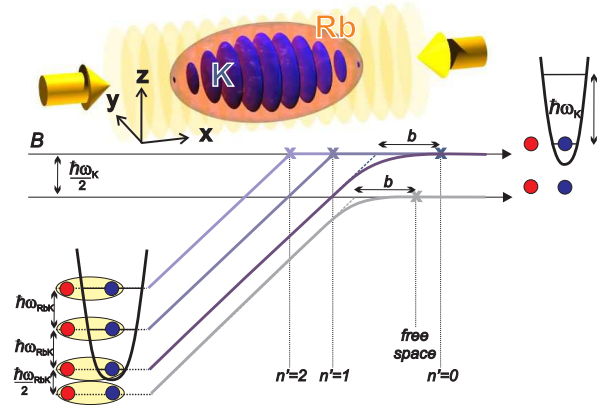}
  \caption{(color online). (top)
Sketch of the experimental configuration: 3D(Rb)--2D(K)
mix-dimensional system obtained using a SSDP lattice. (bottom)
Energy
    diagram of the open and closed channel as a function of the
    Feshbach field $B$: due to the external potential, both channels
    are uplifted and split into many levels, separated by the
    relevant harmonic oscillator quanta $\hbar \omega_i$
    ($i$=K,KRb). MDR may occur when closed-channel levels ($n'$, sloping
    lines) intercept populated open-channel levels (horizontal
    lines). The shift $b$ due to the
    channel coupling is assumed equal to that in free-space for
    $n'$=0, and is neglected for $n'>$0.  }
  \label{fig:sketch}
\end{figure}
In this Letter, we report on the first experimental realization of
a mix-dimensional system composed of two ultracold atomic species,
\kq\ and \rb, of which we control the mutual interactions. We
exploit the technique of species-selective dipole potential (SSDP)
proposed in Refs.\,\cite{PRA07-LeBlanc, PRA06-Massignan} and
implemented in Ref.\,\cite{PRL09-Catani}, to realize a tight
potential confining one species in 2D (\kq), while having a
negligible effect on the other (\rb), that remains 3D. In this
configuration, for the first time we observe a series of up to 5
scattering resonances, induced by the mixed dimensionality. These
discrete resonances are peculiar to configurations having only one
collisional partner tightly confined: they are indiscernible in 3D
homogeneous (or weakly confined) systems and absent for confined
particles with equal harmonic frequencies. Important physical
insight about mix-dimensional resonances (MDRs) can be gained by
the following simple picture. In general, a scattering resonance
may occur when the energy of a closed-channel state equals the
energy of scattering partners. For tightly confined atoms, the
energy shifts of both the scattering threshold (unbound atoms) and
the closed-channel levels (dimer) cannot be neglected. In a
collision between an atom $A$ lying in the harmonic ground level
along the confined direction $x$ and a free atom $B$ with
negligible momentum, resonances might arise when:
\begin{equation}
  \label{eq:energy_crossing_ho}
  \frac{\hbar\omega}{2}=\hbar \omega
 \sqrt{\frac{m_A}{m_A+m_B}}\left(n'+\frac{1}{2}\right) -E_b\quad n'=0,1,2,\dots
\end{equation}
where $\omega$ denotes the harmonic frequency of particle $A$ and
$E_b$ the free-space binding energy of the (bare) closed channel.
Channel coupling, neglected in \eref{eq:energy_crossing_ho}, causes a
resonance position shift $b$, assumed to be relevant only for $n'$=0. The
series of resonances is absent if the colliding atoms have equal
harmonic frequencies because, in this case, the center-of-mass and the
internal motion are decoupled and collisions cannot change the
center-of-mass energy. In the following, a collection of tight
harmonic traps along one direction is provided by a 1D SSDP optical
lattice (\fref{fig:sketch}), such that the harmonic frequency $\omega$
of the confined K sample depends on the lattice strength $V_{\rm
  lat}^K$, hereafter expressed in units of K recoil energy [$V_{\rm
  lat}$=$s (\hbar k_L)^2/(2 m_K)$=$s\cdot h \cdot 7.797$\,kHz].\\
\begin{figure}[t]
  \centering
  \includegraphics[width=0.9\columnwidth]{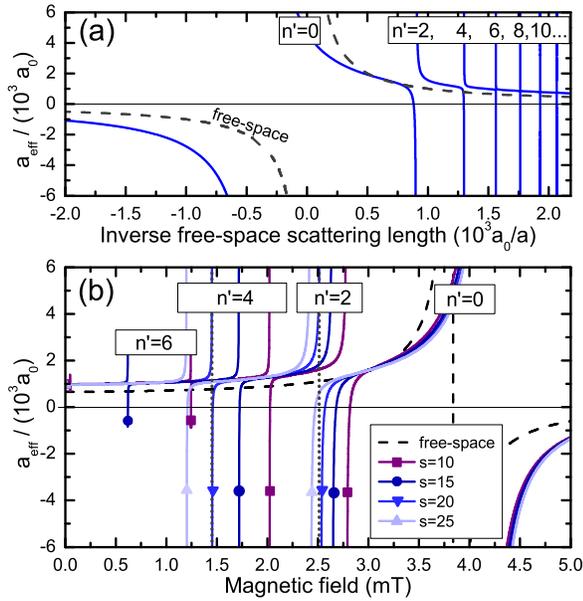}
  \caption{(color online). (a) Calculated $\aeff$
    as a function of $1/a$ for the masses of our mixture and $s$=20. The dashed line
    corresponds to $\aeff$=$a$. Only
    even $n'$ resonances are allowed (see text). (b) Calculated $\aeff$ as a function of the magnetic field,
    for several lattice strengths $s$. The dotted lines show the
    resonance positions calculated from
    \eref{eq:energy_crossing_ho} for $n'=$2,4 and $s$=20.}
  \label{fig:nishida}
\end{figure}
The formal theory of scattering confirms the simple picture above
and allows to derive the scattering amplitude, as well as to
define an effective scattering length $\aeff$. Here we summarize
our analysis following Refs.~\cite{PRA06-Massignan,PRL08-Nishida},
and introducing an effective-range correction that provides an
improved approximation for the binding energy beyond the universal
region $E_b\propto(1/a^2)$, $a$ being the free-space scattering
length.

We consider atoms $A$ and $B$ interacting through a short-range
potential $V(\mathbf r_A$--$\,\mathbf r_B)$. Due to translational
symmetry, we can separate the center-of-mass motion in the $yz$
plane. The Schr\"odinger equation reads:
\begin{eqnarray*}
  &&[H_0+V-E]\psi(x_A,x_B,\bsym\rho_{AB})=0\\
  &&H_0=-\frac{\hbar^2 \partial_{x_A}^2}{2m_A}-
  \frac{\hbar^2\partial_{x_B}^2}{2m_B}
  -\frac{\hbar^2\nabla_{\bsym\rho_{AB}}^2}{2\mu}+\frac{1}{2}m_A\omega^2x_A^2,
\end{eqnarray*}
where $\bsym\rho_{AB}$=$(y_A$--$y_B,z_A$--$z_B)$ and $\mu$ is the
reduced mass.  At short distances
$r$=$\sqrt{(x_A\mathrm{-}x_B)^2+\bsym\rho_{AB}^2}\to0$, the
presence of the confining potential is irrelevant and we can
replace the real potential $V$ with the generalized Bethe-Peierls
boundary condition imposed on the wavefunction:
$\psi|_{r\to0}\propto 1/r$--$1/\tilde a(E_c)$, where $\tilde
a(E_c)^{-1}$=$a^{-1}$--$\mu r_0E_c/\hbar^2$ with $E_c$ denoting
the collision energy \cite{PRL04-Petrov}. On the other hand, the
behavior of the zero energy wavefunction at large distances
$r$$\to$$\infty$ defines an effective scattering length $\aeff$:
$\psi|_{E\to\frac{\hbar\omega}2,r\to\infty} \propto
\left[1/\sqrt{\frac{m_B}\mu x_B^2+\bsym\rho_{AB}^2}-1/\aeff\right]
e^{-x_A^2/(2l_{\rm ho}^2)}$~\cite{PRL08-Nishida}.

The numerically computed $\aeff$ is plotted in
\fref{fig:nishida}(a) for the mass values of our experiment
($A$=\kq, $B$=\rb), the effective range $r_0$=168.37$\,a_0$, and
the confinement length $l_{\rm ho}$=$\sqrt{\hbar/m_A
\omega}$=$1124\,a_0$.  We see that $\aeff$ diverges for one
particular negative value of $a$ and for an infinite series of
positive values of $a$. The width of successive resonances
decreases and, for $l_{\rm ho}/|a|$$\gg$1, $\aeff$ approaches
$a\sqrt{m_B/\mu}$. We also see that the resonant $a$ values
coincide with those calculated by using
\eref{eq:energy_crossing_ho}. The theoretical predictions outlined
in \fref{fig:nishida}, combined with the knowledge of the magnetic
field dependence of the scattering length in free-space, allow for
quantitative predictions of the MDRs positions as well as the
behavior of $\aeff(B)$ [\fref{fig:nishida}(b)].

In free-space, two Feshbach resonances located at 3.84\,mT and 7.87\,mT are
useful to tune the interspecies scattering length
\cite{PRL08-Thalhammer}. The binding energies of the weakly bound
Feshbach \kq \rb\ dimers have been measured \cite{PRA08-Weber} and
used to adjust the collisional model \cite{NJP09-Thalhammer}. For
the weakly bound dimer state arising at 3.84\,mT, the numerically
calculated scattering length as a function of the magnetic field
$B$, in mT, is well reproduced by the parametrization
$a(B)$\,=\,208\,$a_0$\,(1+3.09/($B$+3.852)--\,4.992/($B$--\,3.837)--\,0.164/($B$--\,7.867)),
while the calculated dimer free-space binding energy is well
fitted by the effective range expansion \cite{PRL04-Petrov}:
$E_b(B)=(\hbar^2/\mu
r_0^2)(1$--\,$r_0/a(B)$--\,$\sqrt{1-2r_0/a(B)})$, with
$r_0=168.37\,a_0$ for $B$ ranging from 0 to 3.84\,mT.

In the experiment, we load a crossed dipole trap, created by two
orthogonal, linearly polarized, laser beams ($\lambda$=1064\,nm,
waists$\,\simeq 70\,\mu$m), with a \kq-\rb\ mixture in the
$|F$=$1, m_F$=$1\rangle$ hyperfine state, at 1.5\,$\mu$K. By
lowering the beams intensity, we evaporatively cool the atoms to
$0.3\,\mu$K in presence of a uniform field of 7.7\,mT for which
the interspecies scattering length is convenient
($\sim$$260\,a_0$) to ensure both fast thermalization and a low
rate of inelastic collisions. The final temperature is
sufficiently high to avoid Bose-Einstein condensation of the
samples, but also low enough to make sure that \kq\ occupies only
the ground state of the tight confining potential ($k_B
T$$\ll$$\hbar \omega$). At this point, we ramp the Feshbach field
to 1.4\,mT in 20\,ms and, immediately afterwards, we raise a 1D
SSDP lattice with an exponential ramp of 50~ms (time constant of
10 ms). We then bring the Feshbach field to the final value $B$ in
15~ms and wait typically for 65~ms. We verified that no
discernible fraction of \kq\ atoms lies in excited bands. During
the hold time the atom number decays through inelastic 3-body
recombination collisions; after the lattice is linearly
extinguished in 3\,ms and the atoms expand freely for 6\,ms, we
record the number and temperature of both atomic samples for
different values of the final Feshbach field. The presence of the
scattering resonance is detected as a pronounced peak in the atom
loss due to a resonant enhancement in the 3-body recombination
rate.

The SSDP lattice is a standing wave along the horizontal $x$
direction, with linear polarization oriented along the $z$
direction of the Feshbach field and waist equal to 85\,$\mu$m. The
SSDP wavelength $\lambda=2\pi/k_L$=790.018(2)\,nm is chosen by
minimizing the effect of Raman-Nath diffraction on Rb atoms. With
respect to our previous work \cite{PRL09-Catani}, we improved the
extinction ratio $V_{\rm lat}^{\rm Rb}/V_{\rm
  lat}^{\rm K}$ to be lower than $10^{-2}$ by purifying the lattice polarization.

\begin{figure}[t]
  \centering
  \includegraphics[width=.9\columnwidth]{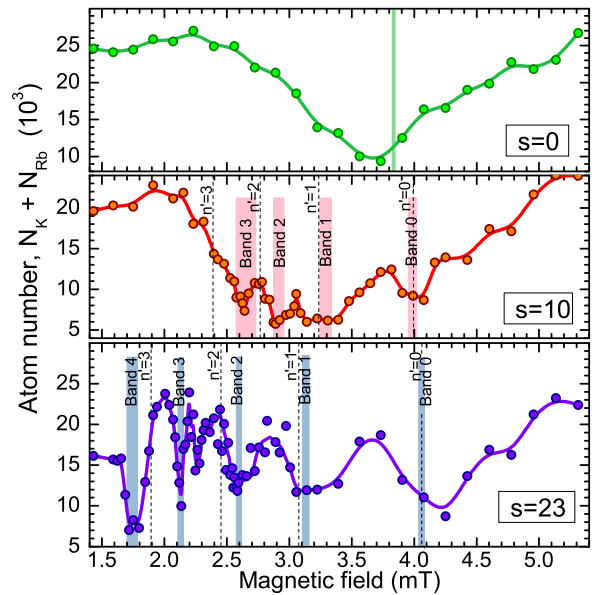}
  \caption{(color online). Recorded atom number $N_{\rm K}$+$N_{\rm Rb}$,
   versus the Feshbach magnetic field without lattice
    (top), with a SSDP lattice strength $s$=10 (middle) and
    $s$=23 (bottom). Solid lines are a guide to the eye.
Predictions based on the harmonic oscillator analysis,
\eref{eq:energy_crossing_ho} (dashed) and on band structure,
\eref{eq:energy_crossing} (shaded areas) are shown.}
  \label{fig:scans}
\end{figure}

\begin{figure}[t!]
\centering
\includegraphics[width=.9\columnwidth]{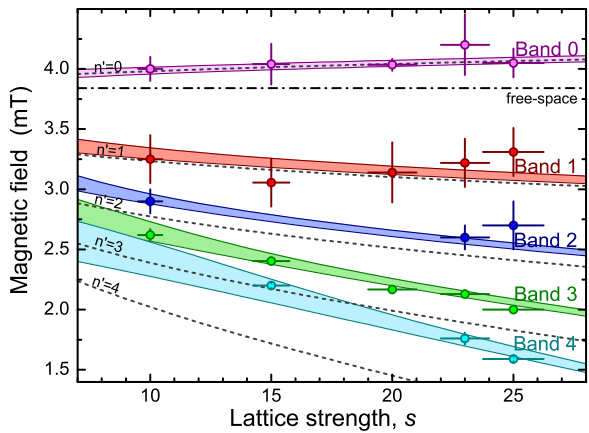}
\caption{(color online). Summary of observed loss peak centers as
a function of the lattice strength $s$. Resonance positions are
extracted from magnetic field scans (see Fig.~2), by fitting the
bottom of the loss peaks with a gaussian. Error bars are dominated
by the width of the loss peaks (vertical) and by the systematic
uncertainty in the lattice calibration (horizontal). Predictions
based on the harmonic oscillator analysis,
\eref{eq:energy_crossing_ho} (dashed) and on band structure,
\eref{eq:energy_crossing} (shaded areas) are shown.}
\label{fig:resume}
\end{figure}

We observe a quite rich spectrum of inelastic losses with several
minima of the total atom number, such that a comprehensive
identification of the different peaks requires broad magnetic
field scans from approximately 1.5 to 5\,mT with a resolution of
7.5\,$\mu$T (\fref{fig:scans}). The first notable feature is that,
already for $s$=10 at the magnetic field of the free-space
Feshbach resonance ($B_0$=3.84\,mT) there is no minimum of atom
number. Instead we find a peak of losses at higher magnetic field,
around 4.0\,mT, where $a<\,$0. Several additional minima are found
in a magnetic field region where $a>\,$0. Position and width of
these peaks depend on the lattice strength: as we increase $s$,
the peaks get narrower and shift towards lower $B$ fields, with
the exception of the minimum at $B>B_0$ which slightly shifts in
the opposite direction.  In \fref{fig:scans}, we also show the
predictions of MDR positions given by the above reported analysis.
While the model correctly describes the trends, the agreement with
the peak positions is qualitative. In order to make more accurate
predictions, we take into account that the lattice actually
creates many adjacent wells that can be treated as pure individual
2D traps for \kq\ only on timescales much shorter than the hopping
time between neighboring sites, $\tau_h$. Since in our case
$\tau_h$ is comparable to the experimental duration, \kq\ atoms
can indeed delocalize over the lattice. As a consequence, we
introduce the energy band structure and calculate the positions of
the expected resonances by means of:
\begin{equation}
  \label{eq:energy_crossing}
  p^2/(2 \mrb)+\epsilon_{\rm K}(0,q;V_{\rm lat}^K)=
\epsilon_{\rm KRb}(n',q+p;V_{\rm lat}^K)-E_b,
\end{equation}
where $\epsilon_i(n,q;V_{\rm lat}^K)$ denotes the energy of the
Bloch wave of particle $i$(=K, KRb), with quasi-momentum $q$ in
the $n$-th band at the lattice potential $V_{\rm lat}^K$, and $p$
the initial momentum of the Rb atom. Like
\eref{eq:energy_crossing_ho}, also \eref{eq:energy_crossing} is
based on two assumptions. First, the binding energy of the dimer
in an excited band equals that in free-space, $E_b$: this
assumption is reasonable whenever the dimer size is smaller than
the on-site oscillator length $(\hbar/\sqrt{\mu E_b}\ll l_{\rm
ho})$, i.e.~outside a small region around the free-space
resonance. Second, we assume the resonance to occur at the energy
crossing, disregarding the shift due to the channel coupling, with
the exception of the $n'$=0 MDR for which we take the same
channel-coupling shift as in free-space ($b\sim 0.45$\,mT).
We expect
that, in presence of the lattice, the channel coupling should be
smaller than in free-space, decreasing as we increase the band
index and the lattice strength.  As shown in \fref{fig:resume},
\eref{eq:energy_crossing} predicts the position of the MDRs with
improved accuracy with respect to the harmonic oscillator
analysis.

In addition, in the case of harmonic confinement, symmetry under
parity inversion allows for coupling only between oscillator
levels differing by an even number of quanta, at zero collision
energy. This is not the case for Bloch waves of generic
quasi-momentum $q$ that are not eigenstates of the parity
inversion operator.  For nonzero collision energy, i.e. $p_{\rm
Rb}$>0, coupling to odd $n'$ harmonic oscillator levels is
allowed, but suppressed at low temperature.  We observe resonances
for all indices $n'$, including odd values, with similar
strengths. In our case, due to the initial temperature
$T$=0.3\,$\mu$K, quite a large part of the first Brillouin zone
($|q|$<$k_L$) is populated by K atoms, since $\sqrt{k_B T
\mk}/\hbar$ $\simeq$$0.62\, k_L$. However, we expect that peaks
with odd $n'$ values are observed for sufficiently long hold times
even at zero temperature, due to the interaction-induced momentum
spread of atoms. For $s$=23 and 25, we also detect a loss peak
located between the $n'$=2 and $n'$=3 peaks. At present we cannot
explain these features, that might be due to few-body physics.

In conclusion, we have realized a binary system, whose components
have different dimensionality, by means of a SSDP lattice that
tightly confines only one of them. By monitoring 3-body inelastic
losses we have observed for the first time a series of multiple
MDRs between (2D) \kq\ and (3D) \rb\ atoms. Approximating the
individual lattice wells as harmonic potentials, a simple argument
based on the degeneracy of open and closed channels explains
qualitatively the behavior of loss peaks and is consistent with
the more complete formal theory. However, the harmonic oscillator
analysis must be replaced with band theory to get a quantitative
agreement with the peaks location and to explain the presence of
odd $n'$ peaks. At present, a comprehensive theory of 3-body
inelastic losses for multiple neighboring MDR is still needed.

Besides their specific interest, mix-dimensional configurations
are also the extreme case of heteronuclear systems with asymmetric
confinement \cite{NJP05-Peano}, that are frequently encountered in
the domain of atomic quantum gases. For weak confinements
($\hbar\omega$$\ll$$E_b$) the effect of the potential is
negligible and its asymmetry is irrelevant, but for tight
confinements, such as those in optical lattice, the scattering
modifications need to be taken into account \cite{PRL04-Fedichev},
as manifested by our findings. Mix-dimensional atomic systems open
many intriguing perspectives: long-lived p-wave trimers close to a
MDR \cite{pwaveTrimers}, a rich Efimov physics
\cite{PRL08-Nishida}, and a new kind of heteronuclear molecules,
whose constituents live in different dimensions, could be
investigated.

This work was supported by MIUR PRIN 2007, Ente CdR in Firenze,
CNR under project EuroQUAM DQS and EU under STREP CHIMONO and
NAME-QUAM. Y.\,N. was supported by MIT Pappalardo Fellowships in
Physics.  We thank M. Jona-Lasinio and the QDG group at LENS for
fruitful discussions.


\end{document}